\documentclass[10pt, conference, compsocconf]{IEEEtran}
\usepackage{cite}
\usepackage[pdftex]{graphicx} \graphicspath{{figs}}
\usepackage[cmex10]{amsmath}
\usepackage[T1]{fontenc}
\usepackage[utf8]{inputenc}
\usepackage{lmodern}
\usepackage{hyperref}
\usepackage{fixltx2e}
\usepackage{stfloats}
\usepackage{url}
\usepackage{color}
\usepackage{amssymb,amsmath,amsfonts}

\begin{document}
\title{Cactus: Issues for Sustainable Simulation Software}

\author{
 \IEEEauthorblockN{
  Frank L\"offler       \IEEEauthorrefmark{1}\IEEEauthorrefmark{6},
  Steven R. Brandt      \IEEEauthorrefmark{1}\IEEEauthorrefmark{2},
  Gabrielle Allen       \IEEEauthorrefmark{3} and
  Erik Schnetter        \IEEEauthorrefmark{4}\IEEEauthorrefmark{5}\IEEEauthorrefmark{1}
 }
 \IEEEauthorblockA{\IEEEauthorrefmark{1}
   Center for Computation and Technology, Louisiana State University, Baton Rouge, Louisiana 70803}
 \IEEEauthorblockA{\IEEEauthorrefmark{2}
   Department of Computer Science, Louisiana State University, Baton Rouge, Louisiana 70803}
 \IEEEauthorblockA{\IEEEauthorrefmark{3}
   Skolkovo Institute of Science \& Technology, Moscow, Russia}
 \IEEEauthorblockA{\IEEEauthorrefmark{4}
   Perimeter Institute for Theoretical Physics, Waterloo, Canada}
 \IEEEauthorblockA{\IEEEauthorrefmark{5}
   Department of Physics, University of Guelph, Guelph, Canada}
 \IEEEauthorblockA{\IEEEauthorrefmark{6}
   knarf@cct.lsu.edu}
}

\maketitle

\begin{abstract}
The Cactus Framework is an open-source, modular, portable programming
environment for the collaborative development and deployment of scientific
applications using high-performance computing. Its roots reach back to 1996 at
the National Center for Supercomputer Applications and the Albert Einstein
Institute in Germany, where its development jumpstarted. Since then, the
Cactus framework has witnessed major changes in hardware infrastructure as well
as its own community. This paper describes its endurance through these past
changes and, drawing upon lessons from its past, also discusses future
challenges for Cactus.
\end{abstract}

\begin{IEEEkeywords}
frameworks, scientific computing, software sustainability
\end{IEEEkeywords}

\IEEEpeerreviewmaketitle

\section{Introduction}
Motivated by the needs of the numerical relativity research community and
stemming from earlier efforts at the National Center for Supercomputer
Applications in the U.S., the design and development of the Cactus
framework~\cite{Cactuscode:web,Goodale:2002a} began at the Albert Einstein
Institute, a Max Planck Institute for Gravitational Physics in 1996. The
component-based architecture of Cactus was inspired by experiences of
physicists and computer scientists who had previously worked together in the
USA Binary Black Hole Alliance Grand Challenge. This NSF-funded collaboration
(1993-1999) involved over eleven groups, working with a variety of independent
code bases on a set of different projects with the aim of modeling the inspiral
collision of two black holes using then-state-of-the-art
supercomputers\footnote{The first accurate black hole inspiral was finally
modeled in 2007.}. Even within a single group, multiple codes were used, often
with multiple versions of each code. Contrary to the spirit of collaboration,
advances in research methods or computing technologies were re-implemented,
debugged, and verified in each code, thus duplicating effort, hampering
communication, and slowing scientific progress.

The vision of the Cactus team was to provide an organic, community-oriented
framework that would allow researchers to easily work together with reusable
and extendable software elements. 

From the beginning, the Cactus framework followed a completely modular design.
It features a comparably small core (named the ``flesh'') which provides the
interfaces between modules at both compile- and run-time. The Cactus modules
(called ``thorns'') use these APIs to specify inter-module dependencies, e.g.
to share or extend configuration information, to use common variables or
run-time parameters. Modules compiled into an executable can remain dormant at
run-time. Based on user-specified parameters and simulation data itself, the
flesh decides when and in which order to call functions in different modules,
assembling them into a coherent simulation.

This usage of modules and a common interface between them enables researchers
to 1) easily use modules written by others without the need to understand all
details of their implementation; 2) write their own modules without the need to
change the source code of other parts of a simulation in the (supported)
programming language of their choice; and 3) easily communicate the
scientifically relevant ideas behind the module without involving the
infrastructural details. The number of active modules within a typical Cactus
simulation ranges from tens to hundreds and often has an extensive set of
inter-module dependencies.

\section{An evolving framework community}
The accelerating growth and diversity of the Cactus community reinforced the
modular development of both the physics-based and computational infrastructure.
Examples of modules include the evolution equations for General Relativity,
radiation or reflective boundary conditions, MPI parallelization, parameter
parsing, and output routines. The value of this modularization is hard to
overstate.

This design was of tremendous help to the motivating science problem, numerical
relativity. It also became clear that with such a design, the modules could
easily be purposed for other science problems. PDE problems are especially
well-suited for use with the Cactus framework, however, since Cactus was
initially written for numerical relativity, which solves a set of complex,
partial differential equations (PDEs). 

The straightforward reusability of existing components spurred development in
other areas of science--usually numerically similar--in particular solving sets
of PDEs, e.g., coastal simulations of storm surges. However, most users of the
Cactus framework by far are interested in numerical relativity, or more
generally, relativistic astrophysics. Cactus development within this field is
coordinated within the Einstein
Toolkit~\cite{Loffler:2011ay,EinsteinToolkit:web}. This focus stems, at in
least part, from the fact that during most of Cactus development, interaction
between developers and users was tight. In fact, most users became developers
to some degree relatively quickly, so that users and developers were never
truly distinct categories. One of the main reasons for this is the modular
plug-in nature of Cactus, allowing end users to add new thorns to their
application.

The majority of developers saw as their main motivation the pursuit of
numerical relativity simulations rather than the broadening of the user base of
the framework itself to other areas of science. However, owing to the modular
nature of the framework, such a broadening was brought about as a result. 

Today the Cactus and the Einstein Toolkit communities are still strongly
interlinked through individuals who are active in both communities; but as
different subdomains expand their usage of Cactus, these two groups polarize
into different roles. Already, most users of the Einstein Toolkit are not
active developers of the underlying framework itself, but rather merely use it
to create and extend modules within the Einstein Toolkit, still occasionally
contributing through new infrastructure; but their use of the framework informs
the infrastructural developers' priorities.

\section{Software Sustainability Issues}
There are many issues connected to the sustainability of software. Some of
these are important for almost all software projects, but some aspects are
especially relevant for scientific projects. Out of the latter the authors
picked four of the most relevant to discuss in more detail.

\subsection{Modular Design}
One of the key properties of a long-term sustainable scientific software is a
modular design. Cactus chose a unique method of enabling modularity that went
beyond the usual notions of APIs, standard data structures, and coding
conventions. Cactus uses a small set of domain specific languages (DSLs) to
describe its distributed data structures and scheduling~\cite{Seidel:2010bb}.
These DSLs enable Cactus modules to do run-time reflection, in both Fortran and
C/C++, on the grid variables being evolved.

This may not sound like a revolutionary idea, but its consequences were
far-reaching. Because of this simple design decision, several things became
possible. First, Cactus was able to completely decouple I/O from science code.
Unlike many scientific codes that have calls to I/O routines interspersed with
the program logic, science modules in Cactus are only concerned with what
variables they read and write. The I/O module(s) can take field variable names
as a parameter, look them up with run-time reflection, and write them out as
text, HDF5, JPEG and other formats. The DSL describing scheduling identifies
when this I/O will be performed, steered at start or even run-time by the user.
Even the most important form of I/O, checkpoint/restart, is enabled and
modularized by this design.

Second, Cactus was able to abstract the time-integration method (e.g.
Runge-Kutta, Iterative Crank-Nicholson, etc.) from the time evolution
equations. The DSL describing scheduling, combined with the list of variables
to be evolved, was sufficient for this task. The value of this module by itself
is significant. The ability to avoid subtle coding bugs or to try out diverse
integrators, without cluttering the codebase, is of great value. This time
integration module can also easily be used to couple separate physics modules,
such as e.g. the Einstein and Hydrodynamics equations.

Third, Cactus was able to create a web browser interface, interrupting the
schedule tree at key places and allowing variables to be inspected and modified
during an execution. This particular form of parameter steering was naturally
enabled by the key module design decisions.

Fourth, an adaptive mesh refinement module named
Carpet~\cite{Schnetter:2003rb,CarpetCode:web} was added. Before Carpet was
available, only uniform, Cartesian grids could be used for spatial
decomposition within Cactus, and a lot of modules expected to get such
rectangular meshes as input. It was possible to integrate Carpet into Cactus
without almost any change to the science modules. Later, a multi-block mesh
capability~\cite{Schnetter:2006pg} was added with similarly little disruption
to existing codes.

There are many other capabilities enabled by the unique modularization
decisions of the Cactus Framework. Other special modules include those for
debugging, e.g. NaNChecker, those enabling unusual IO, e.g. Twitter, generic
grid modules, generic boundary conditions, timers, interfaces to PAPI counters,
analysis modules, etc.

These unique types of interfaces allow different groups to efficiently work on
one common project, towards one common goal, avoiding unnecessary conflicts or
duplication.

\subsection{Growing Collaborative Community}
In contrast to commercial products, academic scientific software like Cactus is
usually developed in a university setting. Most of the actual development work
is performed by graduate students and postdoctoral researchers who are focused
on science. This poses a threat for long-term stability of any project, because
these developers are typically not very interested in contributing to
infrastructure, and frequently leave their research groups after three to four
years, taking all their knowledge and experience with them. New members of
research groups first need to be trained, and while this time can be shortened
by creating respective courses and documentation,
e.g.,~\cite{FL-Allen2011a,FL-Loeffler2011tg,FL-Zilhao:2013aa}, the constant
flux of developers continues to be a struggle.

The infrastructure development problem is handled within the Cactus community
by connecting its development so closely to a specific science problem that
publications that describe both are possible. This is the way many of the
publications using and extending the Einstein Toolkit are written, containing a
description of the new infrastructure while mainly representing a publication
in physics. On the other hand there are examples of pure computational science
publications as well, e.g.,\cite{Hutanu2010a,Seidel2010a,
Seidel2010b,Zebrowski2011bl,Blazewicz2011gb,Blazewicz:2013aa}. These sorts of
projects are enabled by the ability to provide immediate benefits for the
substantial number of physics problems described by the Einstein Toolkit.

The retention problem is handled by the Cactus community automatically, by
creating an enjoyable programming and development experience. Many students
continue to be interested in and develop for Cactus after they move on to new
positions. They are attracted by the ability to leverage the work of other
physicists and computer scientists in the community, to see their code re-used,
and to collaborate on new research. The ability of Cactus to enable
collaboration, based on its unique modular design and the use of open-source
licensing, make it an attractive tool for students to use in their continuing
research. Thus, rather than really leaving, they expand the collaboration. Not
all students continue to use Cactus, but enough are inspired by it to create
the kind of stable contributions needed to maintain the health of the project.

\subsection{Career paths}
The third issue connects to the main motivation of the workforce as well, but
also affects the group leaders, especially if they are young faculty or trying
to become faculty. In contrast to commercial products a main motivation of
developers in academia is \emph{credit}. Credit is obtained through
publications and citations thereof, which are then used to quantify the
scientific impact of an individual's work, forming the basis for future career
plans as well as promotion and tenure for faculty. The most severe problem for
developers in most computational sciences currently is that while most of the
work is done creating hopefully well-written, sustainable software, the
academic success is often exclusively tied to the solution of the scientific
problem the software was designed for. Tasks that from a software engineering
standpoint are essential, e.g., high usability, well-written and updated
documentation, or porting infrastructure to new platforms, are not rewarded
within this system,

While computations of some form entered almost every aspect of academic
research and have been present in many areas already for years, criteria for
tenure positions especially often rate scientific results much higher than the
infrastructural development that was necessary to achieve these results.
Developers of scientific software are often experts across disciplines, and
thus very valuable as a team member, group leader and lecturer. However, the
scope of open tenure positions is often only very limited to a specific
traditional science, resulting in a lack of career opportunities and thus a
lack of student motivation to fully participate in these cross-discipline
activities.

\subsection{Credit}
All the previously mentioned ways to publish assume these papers are cited when
the described software is used by others. While this is often the case, it is
also not uncommon that relevant citations are missing despite the usage of a
specific software package. Enforcement of citation is hard to impossible to
enforce especially for open-source software like the Cactus framework. Any
requirement for citation would conflict with its free-software license.
Changing the license on the other hand, e.g., to include such a requirement, is
also not desired, because there are legitimate cases where such citations are
not possible, e.g., because of space limitations.

Both the Cactus framework and its largest user group, the Einstein Toolkit,
have a few general publications that are requested (but not required) to be
cited whenever the software is used, and especially the Einstein Toolkit also
requests additional citations when some of its modules are used that are so
complex that they are described in publications of their own and deserve extra
credit. Examples of such modules are the Carpet mesh-refinement infrastructure,
or the black hole horizon finder~\cite{Thornburg:2003sf}. Handling citations on
a module-level like this enabled individual developers to receive credit for
their work, especially if they entered the group only after the main
publications had already been written. A complete list of these publications
can be found on the Einstein Toolkit web site~\cite{EinsteinToolkit:web}.

\section{Future Challenges}
Computational science is always evolving, and at an ever increasing pace.
Changes in hardware and progress in numerical methods have been important
factors in the quest to solve increasingly complex problems within a relatively
constant time. The Cactus group is looking for ways to use modularization to
attack these new types of problems.

Prior to the discovery of a stable set of numerical techniques for evolving
black holes, it was necessary to experiment with the form of the Einstein
equations. Unfortunately, these equations contain many hundreds of terms and
transforming them into code was daunting, tedious and error-prone. Sascha Husa
and Ian Hinder developed a tool called Kranc~\cite{Husa:2004ip,Kranc:web} to
mitigate this problem. Kranc takes a tensorial equation written in Mathematica
and generates a complete Cactus module in C++ to evaluate it, enabling
scientists to more easily experiment with the form of equations. Kranc turned
out to be a much more powerful and useful tool than anticipated.

Modern architectures increasingly rely on vectorization to achieve performance,
many of them now performing eight floating point instructions per clock cycle.
Unfortunately, few compilers can vectorize the Einstein equations because of
the sheer number of terms they contain, leading to a large performance penalty.

Recent work shows that Mathematica's pattern matching ability allows us to
transform equations as Kranc generates them, generating explicit vector
instructions and side-stepping compiler limitations. 

Kranc provides another mechanism and opportunity for modularization. Because it
isolates the high level representation of the equations from their
implementation in code, it makes it possible to generate code for multiple
languages (such as CUDA, and OpenCL), and for multiple numerical methods as
well (e.g. Discontinuous Galerkin, finite differencing, etc.).

Thus, equation generation provides a possible path forward for increased
modularization and functionality for the Cactus framework.

\section*{Acknowledgments}
The authors would like to thank Edward Seidel, for his inspiration and vision,
his support and guidance over many years of development of the framework. We
would like to thank the early pioneers of Cactus, including Paul Walker, Joan
Massó, Tom Goodale, and Thomas Radke. More recently, we have to especially
thank developers like Ian Hinder and Roland Haas, whose contributions have been
invaluable for the community. We are also grateful to all the other people who
contributed to Cactus via ideas, code, documentation, and testing; without
these contributions, this framework would not exist today.

Cactus was and is developed with support from a number of different sources,
including support by the US National Science Foundation under the grant numbers
0903973, 0903782, 0904015 (CIGR) and 1212401, 1212426, 1212433, 1212460
(Einstein Toolkit), 0905046, 0941653 (PetaCactus), 1047956 (Eclipse/PTP), a
Deutsche Forschungsgemeinschaft grant SFB/Transregio~7 (Gravitational Wave
Astronomy), and a Canada NSERC grant. 

\bibliographystyle{IEEEtran}
\bibliography{manifest/einsteintoolkit.bib,cactus.bib}
\end{document}